%% file: 13_09_18_CNT.tex
\documentclass[pra,twocolumn,showkeys,showpacs,10pt,superscriptaddress,aps]{revtex4-1}
\usepackage[intlimits]{amsmath}
\usepackage{dcolumn}
\usepackage[english]{babel}
\usepackage{epsfig}
\usepackage{amssymb}
\usepackage{dsfont}
\input{macro.inc}
\usepackage{units}
\graphicspath{{Skizzen/}}

\usepackage{textcomp}

\begin{document}

\newcommand{\affT}{Eberhard-Karls-Universit\"at T\"ubingen, Physikalisches Institut, D-72076, Germany}
\newcommand{\affU}{Universit{\"a}t Ulm, Institut f{\"u}r Quantenphysik and Center for Integrated Quantum Science and Technology (IQ$^{ST}$), D-89069 Ulm, Germany}

\newcommand{\affTUDA}{TU Darmstadt, Institut f\"ur Angewandte Physik, D-64289 Darmstadt, Germany}

\title{Immersing carbon nano-tubes in cold atomic gases}
\author{C. T. Wei\ss} 
\affiliation{\affT}
\affiliation{\affU} 
\author{P. V. Mironova}
\affiliation{\affTUDA}
\author{J. Fort\'agh} 
\affiliation{\affT}
\author{W. P. Schleich}
\affiliation{\affU} 
\author{R. Walser}
\affiliation{\affTUDA}
\email{Reinhold.Walser@physik.tu-darmstadt.de} 
\date{\today}

\begin{abstract}
We investigate the sympathetic relaxation of a free-standing, 
vibrating carbon nano-tube that is mounted on an atom chip and 
is immersed in a cloud of  ultra-cold atoms. Gas atoms colliding 
with the nano-tube excite phonons via a Casimir-Polder
 potential. We use Fermi's Golden Rule to estimate the relaxation 
rates for relevant experimental parameters and develop 
a fully dynamic theory of relaxation for the multi-mode phononic 
field embedded in a thermal atomic reservoir. Based on 
currently available experimental data, we identify the relaxation rates as a function of atom density and temperature that are required for sympathetic ground state cooling of carbon 
nano-tubes. 
\end{abstract}



\maketitle

\section{Introduction}
In 1959, Richard Feynman gave the visionary talk "There's plenty of room at the bottom" \cite{feynman59} and drew the road map for the coming quantum technologies. Half a century later, we are still amidst the silicium era and enjoy many gadgets based on lithographically manufactured nano-electronics. In terms of miniaturization, 
many sectors of nano-technology \cite{rodgers08} have reached the proverbial quantum bottom, but today, 
this mile stone is recognized rather as a resource for novel devices \cite{nielsen00,Aspelmeyer12} than 
the ultimate limit.

Currently, there are many activities 
\cite{milburn12,Aspelmeyer12} to combine 
well characterized individual quantum systems, like trapped ions \cite{Zoller2004,denschlag10}, degenerate quantum gases  \cite{proukakisbook13}, 
super-fluid/superconducting Josephson junctions \cite{grupp13},
quantum dots \cite{recati05}, or nano-mechanical oscillators \cite{wilson04,aspelmeyer11,Poot2012}, with microwave guides, optical resonators or fibers \cite{mekhov12}, to form hybrid quantum systems. Possible applications of such systems range from high-precision force and mass measurements to quantum computation \cite{Kippenberg2012,Machnes2012,Steele2012,Treutlein2010,FortaghBirkl}.

In this context, carbon nano-tubes \cite{smalley97,dresselhaus99} 
are a particularly promising form of rolled-up, mono-layered 
carbon sheets. These families of fullerenes \cite{kroto85} 
combine fascinating aspects of electrical conductivity, 
exceptional mechanical tension moduli with the ability to be 
grown on demand,
or to be deposited mechanically on atomic chips
\cite{Schmiedmayer2002,FortaghScience,FortaghRMP}.
Such features render carbon nano-tubes on atom chips ideal candidates 
to explore quantum mechanical limits
\cite{wilson04,zippilli09,weissbook10,Poot2012}, to construct
single atom detectors \cite{goodsell10} with them,
or to hybridize them with matter waves \cite{schneeweiss12}.

Polarizable particles in front of surfaces are affected by Casimir-Polder potentials \cite{casimir48,Lamoreaux2007,KlimchitskayaRMP}. 
The measurement of such minute forces 
with ultra-cold atoms is of high interest 
\cite{antezza04,harber05,obrecht07,jetter12}.
In particular, the interaction and trapping of atoms in front of nano-tubes has been explored 
\cite{Scheel07,Klimchitskaya08}. 
However, so far only a few experiments have been 
performed with carbon-nano tubes immersed in ultra-cold alkali gases \cite{gierling11,schneeweiss12}.

In this article, we will combine aspects of 
the quantum motion of carbon nano-tubes as well as the 
quantum degeneracy of the matter waves 
to examine the prospects for the sympathetic cooling due to the Casimir-Polder interaction of a free-standing carbon nano-tube 
mounted on an atom-chip.
In Sec.~\ref{CNTinAtoms}, we 
discuss a micro-mechanical model of the oscillating carbon 
nano-tube and quantize the phononic excitations.
We introduce the energies of the atomic 
gas, phonons, and atom-phonon interaction. 
Within the Heisenberg picture, we model the temporal evolution and obtain equilibrium relaxation rates versus the temperature for different densities of the atomic cloud in Sec.~\ref{DampRate}. 
This approach is generalized in Sec.~\ref{DensM}, 
where we formulate a fully dynamical theory for the density
operator of the phonons. We conclude the 
theoretical analysis with a numerical study of the cooling 
efficiency of a carbon nano-tube due to the interaction with the 
cold atoms. Two short appendices summarize properties of the carbon nano-tubes and thermodynamic correlation functions for degenerate bosonic gases.

\section{Atoms hitting carbon nano-tubes}
\label{CNTinAtoms}

In the present section, we will introduce the basic mechanical features
of single-walled carbon nano-tubes. In particular, we will consider the experimental setup of P.~Schneeweiss 
{\it et al.} \cite{schneeweiss12}, 
where  free-standing carbon nano-tubes are grown on an atomic chip \cite{FortaghRMP,Schmiedmayer2002}
and interact with ultra cold-atoms in ultra-high vacuum systems. 
 
From a continuum model of a free-standing carbon nano-tube, we 
derive a quantized description of the phononic excitations. This 
carbon nano-tube is embedded in an ultra-cold atomic bosonic gas 
at temperatures above the critical temperature for Bose-Einstein 
condensation (BEC).
In particular, we model 
the interaction between a carbon nano-tube and a $^{87}$Rb alkali 
atom by a Casimir-Polder interaction potential 
\cite{schneeweiss12}. 
This approach leads to a total energy for the  carbon nano-tube 
immersed in an atomic bath.

\subsection{Vibrating carbon nano-tube}

The vibrations of a carbon nano-tube can be well described by the Euler-Bernoulli model of an oscillating beam 
\cite{Landau87,Poot2012,smalley97,Ruoff2002}. Here, we use a real, transversal displacement field   $\mathbf{u}(z,t)\equiv u_x(z,t)\mathbf{e}_x+u_y(z,t)\mathbf{e}_y$ to represent the two-dimensional bending of an elastic beam of the length $L$, which is aligned along the surface-normal ($z$-axis) of an atomic chip. The transverse polarization directions are denoted by $\mathbf{e}_{x}$ and 
 $\mathbf{e}_{y}$. 
To be precise, we have to specify the boundary conditions 
$\mathbf{u}(0,t)=\partial_z \mathbf{u}(0,t)=0$, for the fixed end on
 the chip 
and $\partial^2_z \mathbf{u}(L,t)=\partial^3_z \mathbf{u}(L,t)=0$
 on the loose end.

In the case of small displacements, the vibration  of the 
tube follows from the linear  Euler-Bernoulli equation 
\begin{equation}
\label{EBEqn}
 (\rho_c\,{\partial_t^2}+EI\,{ \partial_z^4})\mathbf{u}(z,t)=0.
\end{equation}
The physical parameters  are 
\cite{Ruoff2002,Benaroya1999,Bachtold2007} the linear mass 
density $\rho_c$ [kg/m], the Young's modulus $E$ [Pa] and the 
area moment of inertia $I$ [m$^4$].

The general solution of \eq{EBEqn} can be written as 
\begin{equation}\label{u}
\mathbf{u}(z,t)=\frac{1}{\sqrt{2}}\sum\limits_{\substack{l=0\\\sigma=x,y}}^{\infty}
\mathbf{e}_{\sigma}\phi_l(z)(e^{-i \omega_l t}\beta_{l\sigma}
+e^{i \omega_l t} \beta_{l\sigma}^\ast),
\end{equation}
if we can determine the eigenfrequencies 
$\omega_l$ and  
the phononic eigenmodes $\phi_l(z)$ of the problem. 
From the axial symmetry of \eq{EBEqn}, it follows that the transverse oscillation frequencies must be degenerate for both polarization directions.
The complex amplitudes $\beta_{l\sigma}$ are determined from the initial conditions. 

This model of an oscillating beam, together with the boundary 
conditions leads to a self-adjoint eigenvalue problem for the 
modes with respect to the real scalar product 
\begin{equation}\label{orthogonality}
 \braket{\phi_l}{\phi_m}\equiv
\int\limits_0^L \frac{\rmd z}{L}\phi_l(z)\phi_m(z)=
a_l^2\delta_{l,m}.
\end{equation}
Explicitly, the eigenmodes read
\begin{align} \nonumber
\phi_l(z)=&\tilde{a}_l[(\cos{\kappa_l L}+\cosh{\kappa_l L})
(\cos{\kappa_l} z-\cosh{\kappa_l z})\\
 &+(\sin{\kappa_l L}+\sinh{\kappa_l L})(\sin{\kappa_l z}
-\sinh{\kappa_l z})],
 \label{eigenmode}
\end{align}
where the normalization constants $\tilde{a}_l$, defined 
by \eq{eqnorm}, are
proportional to the harmonic oscillator length 
$a_l\equiv \sqrt{\hbar/\omega_l\rho_c L}$.

However, a mechanical oscillation is only possible, if the wave numbers $\kappa_l$ satisfy the condition
\begin{equation}
\cos{(\kappa_l L)}\cosh{(\kappa_lL)}=-1.
\end{equation}
At first glance, the solution of this equation looks like a formidable task, but in fact 
the phononic wave numbers can be approximated quite well 
\cite{weissbook10}
by an equidistant array $\kappa_l\cong\pi(l+1/{2})/L$, 
for $l\ge 0$.

From the solution of \eq{EBEqn}, we obtain finally the  particle-like dispersion relation 
\begin{equation}
 \omega_l=\sqrt{\frac{EI}{\rho_c}}\kappa_l^2,
\end{equation}
for the phononic frequency $\omega_l$ of mode  $l$.
Such a  quadratic phonon dispersion relation  can be derived also microscopically from the zone-folding method~\cite{dresselhaus00}, considering the  lattice symmetries of graphene~\cite{weissbook10}. 

Having determined the eigenmodes of the tube, we represent the energy of the phonon field 
\begin{align}
\label{H_b}
H_c&\equiv\int\limits_0^L\rmd z\,
\left[\frac{\rho_c}{2} 
\left(\partial_t \mathbf{u}\right)^2+\frac{EI}{2}\,\left(\partial_z^2 \mathbf{u}\right)^2\right]\nonumber\\
&=\sum_{\substack{l=0\\ \sigma=x,y}}^{\infty}
\hbar \omega_l{\beta}_{l\sigma}^{*}{\beta}_{l\sigma}^{\phantom{\dagger}},
\end{align}
as a separable sum of independent oscillator modes, which 
is necessary for the quantization of the phononic field 
\begin{equation}
\label{u_quant}
\hat{\mathbf{u}}(z)=\frac{1}{\sqrt{2}}
\sum_{\substack{l=0\\\sigma=x,y}}^{\infty}
\mathbf{e}_{\sigma}\phi_l(z)(\hat{b}_{l\sigma}^{\phantom{\dagger}}+
\hat{b}_{l\sigma}^{\dagger}),
 \end{equation}
in terms of bosonic excitations
$ [\hat{b}_{l\sigma}^{\phantom{\dagger}},\hat{b}_{l'\sigma'}^{\dagger}]=\delta_{ll'}\delta_{\sigma\sigma'}$,
acting on the multi-mode phononic Fock-states 
$|\ldots,n_{l\sigma},\ldots\rangle$.

Thus, we obtain the Hamiltonian operator
\begin{equation}
\label{hc}
\hat{H}_{c}=\sum_{\substack{l=0\\\sigma=x,y}}^{\infty}
\hbar \omega_l\hat{b}_{l\sigma}^{\dagger}
\hat{b}_{l\sigma}^{\phantom{\dagger}}
\end{equation}
of the vibrating carbon nano-tube.

\subsection{Cold atomic gas}

Atoms colliding with the carbon nano-tube are modeled as a homogeneous 
gas of scalar bosons represented in a discrete 
plane-wave basis as
$
[\aop{\mathbf{k}},\aopd{\mathbf{k'}}]=\delta_{\mathbf{k}\mathbf{k'}}
$.
This representation implies periodic boundary conditions for 
all spatial fields, e.\thinspace{}g., 
the atomic field
$\hat{\Psi}(\br+\mathbf{e}_i L_i)=\hat{\Psi}(\br)$, 
where $L_x$, $L_y$ and $L_z$ are the lengths of the quantization box.

Normalized plane-waves 
\begin{math}
\braket{\br}{\mathbf{k}}\equiv \exp{(i \bk \br)}/\sqrt{\mathcal{V}},
\end{math}
are then a complete set of basis functions within the quantization volume 
$\mathcal{V}\equiv L_x L_y L_z$. 
For the length of the quantization volume in the $z-$direction, we choose the length $L_z=L$ of the carbon nano-tube.

This spatial periodicity leads to a discrete set of atomic wave vectors 
$k_i=2\pi n_i/L_i$, $n_i\in  \mathds{Z}$.
Hence, we obtain for the spatial amplitude of the 
atomic field the expression
\begin{equation} 
\hat{\Psi}(\mathbf{r})=\sum\limits_\mathbf{k}\langle\mathbf{r}|\mathbf{k}\rangle\hat{a}_\mathbf{k},
\end{equation}
and the density of the atomic gas follows as 
\begin{equation} 
\label{density}
\hat{n}(\mathbf{r})\equiv
\hat{\Psi}^\dagger(\mathbf{r})\hat{\Psi}(\mathbf{r})=
\sum\limits_{\mathbf{k},\bq}
\hat{a}^\dagger_{\bq}
\hat{a}^{\phantom{\dagger}}_\mathbf{k}
\braket{\br}{\bk}\braket{\bq}{\br}.
\end{equation}

In this article, we consider atomic temperatures above the 
Bose-Einstein condensation temperature, 
where corrections to the ideal gas energy are minute. Therefore, 
we only keep in the energy the kinetic energy 
\begin{align} 
\label{varepsk}
\hat{H}_{\text{a}}&\equiv \sum\limits_{\mathbf{k}}
\varepsilon_{\mathbf{k}}\hat{a}_{\mathbf{k}}^{\dagger}
\hat{a}^{\phantom{\dagger}}_{\mathbf{k}},
\quad\varepsilon_{\mathbf{k}}\equiv \frac{\hbar^2\mathbf{k}^2}{2m}=\hbar\omega_\mathbf{k}.
\end{align}
of the gas.
\subsection{Casimir-Polder potential}

The interaction of a neutral, polarizable atom in front of  
a surface is described by the Casimir-Polder theory  
\cite{Lamoreaux2007,KlimchitskayaRMP,Klimchitskaya2006,Klimchitskaya2007,Scheel07,judd11}. 
In principle, a full three-dimensional geometric modeling of the 
Casimir-Polder potential surface would be in place to 
describe all details of the interaction. 
However, we assume for simplicity that the dominant 
contribution to the dynamics is obtained by the axis-symmetric, 
translationally invariant part of the Casimir-Polder potential 
$V(\mathbf{r})=V(\rho)$, where $\rho\equiv \sqrt{x^2+y^2}$ is the 
radius in cylindrical coordinates with the $z$-axis oriented 
along the tube. 

Recent experiments \cite{schneeweiss12,jetter12}, have provided evidence 
that the dominant contribution is given by an inverse power-law 
term $\sim C_5/\rho^5$. However, in order to keep the model 
flexible, we approximate the Casimir-Polder potential as an 
inverse 
power series 
\begin{equation}\label{CP}
 V(\rho)=\sum\limits_{n=n_{m}}^{\infty}\frac{C_n}{\rho^{n}},
\end{equation}
for distances well above the tube's 
physical radius $\rho> R$. For example, in case of a 
single-walled carbon nano-tube $R=1$ nm.

The different contributions to the short-range part of the 
potential are parameterized by real coefficients $C_n$, starting  at 
least with terms beyond  $n_{m}>2$. The surface physics 
(adhesion/adsorption) at $\rho=R$ is unknown but should be of no concern
for the calculation of the collisional relaxation. Thus, we 
assume that the potential is real and vanishes within the tube $V(
\rho\le R)=0$.

The later analysis requires the Fourier-transform 
\begin{equation}
\label{V_q} 
V_{\mathbf{q}}\equiv 
\int\limits_\mathcal{V}\rmd^3{r}\,
\braket{\bq}{\br} 
V(\mathbf{r})=
\frac{\delta_{q_z 0}}{\sqrt{\mathcal{V}}}L \,V(q) 
\end{equation}
of the 
 Casimir-Polder potential,
where $V(q)$ is the two-dimensional, axis-symmetric 
Fourier-transform of the Casimir-Polder-potential. 

If the size of the periodic box is much larger than the finite 
range of the potential, one can simply extend the integration limits to the full plane 
\begin{align}
V(q)&=
\sum\limits_{n=n_{m}}^{\infty}C_n
 \int\limits_R^\infty\int\limits_0^{2\pi}
\rmd\rho\rmd\varphi\,
\frac{e^{-i q\rho\cos\varphi}}{\rho^{n-1}}\nonumber\\
&=2\pi\sum\limits_{n=n_{m}}^{\infty}\frac{C_n}{R^{n-2}}V_n(qR).
 \label{Vvonq}
\end{align}
With defining a dimensionless wave number $\bar{q}=qR$, 
the expression 
\begin{equation}
 \begin{split}
{V}_n(\bar{q})=\frac{_1{\rm F}\!_2(1-\tfrac{n}{2};\{1,2-\tfrac{n}{2}\};
-\tfrac{\bar{q}^2}{4})}{n-2} \, 
-\frac{n \Gamma(-\frac{n}{2})}{2^{n}\Gamma(\frac{n}{2})}{\bar{q}}^{n-2}
\end{split}
\label{Vnq}
\end{equation}
for the partial potential amplitudes is formed with the 
generalized hyper-geometric function ${_1}{\rm F}\!_2\left(a_1;\{b_
1,b_2\};z\right)$,
 as well as the gamma function $\Gamma(z)$ \cite{gradsteyn00}.
\begin{figure}[h!]
\centering\includegraphics[clip,width=\columnwidth]{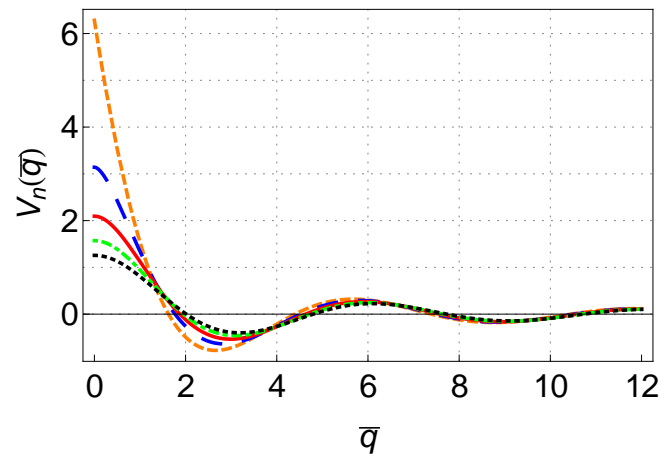}
\caption[Fourier amplitudes of Casimir-Polder potential]{[Color online]:
Dimensionless Fourier amplitudes ${V}_n(\bar{q})$ of the
 Casimir-Polder potential vs. the dimensionless wave number $\bar{q}$ for different inverse power laws: 
$n=3$ (dashed orange), $n=4$ (large-dashed blue), $n=5$ 
(solid red), $n=6$ (dot-dashed green), and $n=7$ (dotted black).}
\label{fig:Vnq}
\end{figure}

For each positive integer $n$ this function can be represented
explicitly in terms of Bessel functions. 
In Fig.~\ref{fig:Vnq}, we depict the behavior of the 
potential amplitude ${V}_n(\bar{q})$ 
in its dependence on the dimensionless momentum $\bar{q}$ for 
the powers $n=3 - 7$. We notice the 
typical oscillatory Bessel-like behavior  with a weak 
dependence on n.

In order to obtain the interaction energy between atoms and the tube, we have to sum the Casimir-Polder potential weighted by the atomic density over all space and over the whole length of the tube leading to the expression
\begin{equation}
\label{Hi1}
 \hat{H}_{\text{i}}\equiv
\int\limits_0^L\frac{{\rm d}z'}{L}
\int\limits_{\mathcal{V}}{\rmd}^3{r}\:\hat{n}(\mathbf{r})V\left[\hat{\mathbf{u}}(z')-\mathbf{r}\right].
\end{equation}

We note that $\hat{H}_{\text{i}}$ is given by a convolution 
integral. For computational purposes, it is more convenient to transform it into the Fourier domain where it becomes a separable sum of products of atomic and tube variables.
Indeed, by introducing the atomic density fluctuations $\hat{\mathcal{N}}_{\bq}$ as the 
Fourier transform 
\begin{align}
\hat{\mathcal{N}}_{\bq}&\equiv
\sqrt{\mathcal{V}}\int_\mathcal{V} \rmd^3 r 
\braket{\bq}{\br} \hat{n}(\br)= 
\sum\limits_{\bk} \hat{a}_{\bk-\bq}^{\dagger}
\hat{a}^{\phantom{\dagger}}_{\bk}=
\hat{\mathcal{N}}^\dagger_{-\bq},
\end{align}
of the atomic density
and a translation operator
\begin{align}
\hat{\mathcal{T}}_\bq&\equiv
\int\limits_0^L\frac{{\rm{d}}z}{L}
\exp[i\bq\cdot\hat{\mathbf{u}}(z)],
\label{Tq}
\end{align}
 of the tube,
we can express the interaction energy as 
\begin{equation}
\label{Hi} 
\hat{H}_{\text{i}}\equiv
\hbar\sum_\bq 
w_\bq
\hat{\mathcal{N}}_{\bq}
\hat{\mathcal{T}}_\bq.
\end{equation}
Here, we have defined the frequency 
$w_\bq\equiv V_\bq/\hbar \sqrt{\mathcal{V}}$ 
to  measure the strength of the Casimir-Polder potential.
\subsection{Total Energy}
\label{sectoteng}
Therefore, the total Hamiltonian 
\begin{equation}\label{H1}
 \hat{H}\equiv \hat{H}_{c}+\hat{H}_{a}+\hat{H}_{\text{i}}
\end{equation}
of the system 
is the sum of the phonon energy $\hat{H}_{c}$, 
the atomic kinetic energy $\hat{H}_{a}$, and the atom-phonon interaction energy $\hat{H}_i$.

Recalling the interaction energy \eq{Hi}, we notice
that the homogeneous background contribution at $\mathbf{q}=0$, depends only on the atomic particle number 
$\hat{N}\equiv\hat{\mathcal{N}}_0$ . Hence, it is convenient
to shift the individual energies by this amount, that is
\begin{eqnarray}
\label{hiprime}
\hat{H}_{\text{i}}&\rightarrow& \hat{H}_{\text{i}}-
\hbar w_0
\hat{N}
=\sum_{\mathbf{q}\neq0}
\hbar w_\bq
\hat{\mathcal{N}}_{\bq}
\hat{\mathcal{T}}_\bq,\\
 \hat{H}_{\text{a}}&\rightarrow& \hat{H}_{\text{a}}+\hbar w_0
\hat{N}=
 \sum\limits_{\mathbf{k}} (\varepsilon_{\mathbf{k}}+\hbar w_0)
 \hat{a}_{\mathbf{k}}^{\dagger}\hat{a}^{\phantom{\dagger}}_{\mathbf{k}}.
\end{eqnarray}
This energy renormalization amounts to dropping 
the term $\bq=0$ in the interaction and to off-set the atomic
 dispersion relation by $\varepsilon_{\mathbf{k}}\rightarrow \varepsilon_{\mathbf{k}}+\hbar w_0$.

Experimentally relevant temperature scales are given by the 
cryogenically cooled atom chip, where the carbon nano-tube is mounted
 on, corresponding to $T_c=4$~K, and by the temperature 
$T_a=100$ nK of the atomic gas.

In Appendix \ref{normalization} and Ref.~\cite{weissbook10}, 
the maximal thermal displacement $u$ of the tube as a function of temperature is estimated. Clearly, $u$ decreases with decreasing the temperature of the chip. 
On the other hand, the thermal de~Broglie wavelength 
$\lambda_{\text{dB}}$ of the atomic gas increases at lower
 temperatures. Hence, if we are in a regime where 
$u\ll \lambda_{\text{dB}}$, we can approximate the tube's displacement operator 
\begin{gather} 
\label{Texpanded}
\hat{\mathcal{T}}_\mathbf{q}
=1+
\int\limits_0^L\frac{{\rm{d}}z}{L}
\{i \mathbf{q}\cdot\hat{\mathbf{u}}(z)-
\frac{1}{2}[\mathbf{q}\cdot\hat{\mathbf{u}}(z)]^2+\ldots\}.
\end{gather}
by the first two terms of the Taylor series of \eq{Tq}.
\section{Static relaxation rate}\label{DampRate}
We are now ready to proceed with the calculation of the static relaxation rate using first-order time-dependent perturbation theory. From the solution of the Heisenberg equations for the phononic fields, we can derive the time-dependent occupation number of the lowest phononic mode. 
Then, we use Fermi's Golden
Rule 
\footnote{Correctly speaking, Fermi's Golden Rule should 
be attributed to P.~A.~M. Dirac, Proc. Roy. Soc. A, {\bf 114}, 
243 (1927); see T.~D. Visser, Am. J. Phys., {\bf 77}, 487 (2009)},
 to calculate the relaxation rate as a function of the temperature
 of the atomic gas. Our analysis demonstrates the possibility of carbon
 nano-tube cooling  in the experiment.

\subsection{Heisenberg equations of motion}
By evaluating the Heisenberg equations of motion
for the atomic and phononic fields, \ie,
$i\hbar \dot{\hat{A}}(t)=[\hat{A},\hat{H}]$,
we obtain 
\begin{align}  
\nonumber
  \dot{\hat{a}}_\mathbf{k}&= 
  -i\omega_\mathbf{k}\hat{a}_\mathbf{k}+
  \sum\limits_{\mathbf{q}\neq0}\sum\limits_{\sigma=x,y}
   w_\mathbf{q} \,\mathbf{q}\cdot\mathbf{e}_\sigma\,
 \hat{a}_{\mathbf{k}+\mathbf{q}}
  \\ \nonumber
&\times\sum\limits_{l=0}^\infty 
\{ I_l (\hat{b}_{l\sigma}^{\phantom{\dagger}}+
\hat{b}_{l\sigma}^\dagger)+i J_l
\sum\limits_{\sigma'=x,y}
\mathbf{q}
\cdot\mathbf{e}_{\sigma'}\\ 
\label{Ha1}
&\times
(\hat{b}_{l\sigma}^{\phantom{\dagger}}
\hat{b}_{l\sigma'}^{\phantom{\dagger}}+
\hat{b}^\dagger_{l\sigma}
\hat{b}_{l\sigma'}^{\phantom{\dagger}}+
\hat{b}_{l\sigma}^{\phantom{\dagger}}\hat{b}_{l\sigma'}^\dagger+
\hat{b}^\dagger_{l\sigma}\hat{b}_{l\sigma'}^\dagger
)\}
\end{align}
and
\begin{align}
\dot{\hat{b}}_{l\sigma}^{\phantom{\dagger}}&=
-i\omega_l\hat{b}_{l\sigma}^{\phantom{\dagger}}+
\sum_{\mathbf{q}\neq0}
 w_\mathbf{q} \,
\mathbf{q}\cdot\mathbf{e}_\sigma\,
\hat{\mathcal{N}}_{\mathbf{q}}\nonumber
\\
\label{Hb1}
&\times
\{I_l+2 i J_l
\sum\limits_{\sigma'=x,y}
\mathbf{q}\cdot\mathbf{e}_{\sigma'}
(\hat{b}_{l\sigma'}^{\phantom{\dagger}}+
\hat{b}_{l\sigma'}^\dagger)\},
\end{align}
where the constants 
$
I_l\equiv
\int\limits_0^L {\rm d} z \phi_l(z)/\sqrt{2}L$ and 
$
J_l\equiv a_l^2/4
$
reflect the spatial extent of the l-th phononic mode.

From the formal solution of \eq{Ha1} for the atomic variables, we obtain
\begin{gather}
\label{freef}
  \hat{a}_\mathbf{k}(t)=
  e^{-i\omega_\mathbf{k}t}\hat{a}_\mathbf{k}(0)
  +\mathcal{O}(w),
\end{gather}
that is a free evolution of the atomic field plus corrections.

First, we are only considering the two lowest degenerate phononic
 modes with $l=0$.
Excited-state modes with $l>0$ will contribute little to the 
collisionally induced excitation rate considering 
the principle of energy conservation.

The two ground-state modes can be combined into a vector 
\begin{equation} 
\hat{\mathbf{b}}_0(t)
\equiv\sum\limits_{\sigma=x,y}\hat{b}_{0\sigma}(t)\mathbf{e}_{\sigma}.
\end{equation}
Within the rotating-wave approximation, we drop the 
counter-rotating term in \eq{Hb1} and obtain a 
Langevin equation 
\begin{equation}
\label{Eq_b}
\dot{{\hat{\mathbf{b}}}}_0=
-i[\omega_0-\hat{\Omega}(t)]
 \hat{\mathbf{b}}_0+\hat{\mathbf{F}}(t),
\end{equation}
for the phononic Kubo-oscillators 
with a hermitian frequency tensor
\begin{align}
\hat{\Omega}(t)&=
2J_0 \sum_{\mathbf{q}\neq0}
w_\bq
(\mathbf{q}\otimes\mathbf{q})_{xy}\,
\hat{\mathcal{N}}_\mathbf{q}(t),
\end{align}
 which results in pressure-broadened oscillation frequencies,
and a stochastic force
\begin{align}
\hat{\mathbf{F}}(t)&\equiv I_0 \sum_{\mathbf{q}\neq0}
 w_\bq
 \mathbf{q}\hat{\mathcal{N}}_\mathbf{q}(t).
\end{align}
The linear inhomogeneous Kubo equation (\ref{Eq_b}), can be solved formally 
as
 \begin{gather}
\label{persol}
\hat{\mathbf{b}}_0(t)=
\hat{U}_{\Omega}(t,0)
\hat{\mathbf{b}}_0(0)+
\int\limits_0^t{\rm d}t_1
\hat{U}_{\Omega} (t,t_1)
\hat{\mathbf{F}}(t_1),
\end{gather}
by introducing the retarded propagator 
 \begin{gather}
\partial_t\hat{U}_{\Omega}(t,t_1)=
-i[\omega_0-\hat{\Omega}(t)]\hat{U}_{\Omega}(t,t_1),
\end{gather}
for times $t\geq t_1$ with the initial condition 
$\hat{U}_\Omega(t,t)=\mathds{1}$.
Thus, we have obtained a linear-response model for the carbon nano-tube immersed in an atomic gas.
\subsection{Occupation number}
In the present subsection, we assume that the state $\hat{\rho}_{\text{tot}}\equiv\hat{\rho}_a
\otimes
\hat{\rho}_c$
of atoms and phonons is  initially uncorrelated.
Furthermore, we consider atoms that are prepared 
in a grand canonical ensemble $\hat\rho_a$ at a 
temperature $T_a\equiv 1/k_B \beta_a$ without macroscopic motion. Then, 
the number of the atoms with wave number $\mathbf{k}$ is given by
\begin{equation}
\label{GrCan}
n_\mathbf{k}\equiv
\langle\aopd{\bk}\aop{\bk}\rangle=
\frac{1}{e^{\beta_a(\varepsilon_k-\mu)}-1},
\end{equation}
where we have introduced the chemical potential $\mu$, or 
alternatively the fugacity $\eta\equiv e^{\beta{_a\mu}}$, to 
constrain the particle number to $N$ as shown in 
App.~\ref{app:fugacity}. 

Moreover, the tube is prepared in its phononic vacuum
$\hat{\rho}_c\equiv \ket{0,0}\bra{0,0}$, representing the 
ground-state for the $x-$ and $y-$polarization of the $l=0$ modes.

As an observable, we consider the unpolarized occupation 
\begin{equation}
\label{n}
p_{0}^{v}(t)\equiv
\av{\mathbf{b}^\dagger_0(t)
\hat{\mathbf{b}}_0(t)}=
\text{Tr}\left\{
\hat{\mathbf{b}}^\dagger_0(t)
\hat{\mathbf{b}}_0(t)
\hat{\rho}_{\text{tot}}\right\}
\end{equation} 
of the degenerate ground-state manifold. 
From \eq{persol}, it follows that the occupation 
\begin{gather}
p_{0}^{v}(t)=
\int_0^t 
{\rm d}t_1 
\int_0^t 
{\rm d}t_2
\av{
\hat{\mathbf{F}}^\dagger(t_1)
\hat{U}_{\Omega}^\dagger (t,t_1)
\hat{U}_{\Omega} ^{\phantom\dagger }(t,t_2)
\hat{\mathbf{F}}(t_2)
}\nonumber\\
=\int_0^t {\rm d}t_1 
\int_0^t 
{\rm d}t_2
e^{i\omega_0(t_2-t_1)}
\av{
\hat{\mathbf{F}}^\dagger(t_1)
\hat{\mathbf{F}}(t_2)}+\mathcal{O}(w^3)
\label{ptt}
\end{gather} 
is determined solely by the force-correlation function.
In the evaluation of \eq{ptt}, we have used the Born approximation keeping only processes up to second order in the Casimir-Polder potential amplitude $w_\bq$.
In turn, the force-correlation function depends fundamentally on 
the density-correlation function 
$\av{\hat{\mathcal{N}}_{\bq}(t_1)
\hat{\mathcal{N}}_{\bk}(t_2)}$, which by means of space-time Fourier-transformation is linked to the dynamic structure factor 
\cite{pines63}. 
For the free fields of \eq{freef}, we have evaluated this finite temperature correlation function in App.~\ref{ttc}.

After performing the time integration, we obtain for the 
occupation the second-order perturbative result
\begin{align}
p_0^{v}(t)&=4I_0^2
\sum_{\mathbf{q}\neq0,\mathbf{k}}
q_\perp^2|w_\mathbf{q}|^2
n_\mathbf{k}
(1+n_{\mathbf{k}-\mathbf{q}})
\frac{\sin^2[\Delta_{\bq\bk} \frac{t}{2}]}{\Delta_{\bq\bk}^2}.
\label{nsum}
\end{align}
Here, we have introduced the wave vector $\mathbf{q}_\perp$ 
perpendicular to the tube axis by the relation
\begin{math}
q_\perp^2=q_x^2+q_y^2
\end{math}, 
and the frequency difference $\Delta_{\bq\bk}\equiv\omega_0+\omega_{\mathbf{k}-\mathbf{q}}-\omega_{\mathbf{k}}$ due to inelastic scattering  between the initial and the final state.

Assuming that the temperature $T_a$ of the gas is above the 
temperature $T_{\text{BEC}}$ for Bose-Einstein condensation, 
that is $T_a>T_{\text{BEC}}$, the thermal occupation 
satisfies the inequality $n_{\mathbf{k}-\mathbf{q}}<1$. 
Consequently, we can disregard this bosonic enhancement of scattering in \eq{nsum} and retain only
\begin{align}
\label{nsum1}
p_0^{v}(t)&\approx
4I_0^2
\sum_{\mathbf{q}\neq0,\mathbf{k}}
q_\perp^2
|w_\mathbf{q}|^2
n_{\bk}
\frac{\sin^2\left[\Delta_{\bq\bk} 
\frac{t}{2}\right]}{\Delta_{\bq\bk}^2}.
\end{align}
Now, we evaluate the occupation number in the continuum limit 
\begin{math}
 \sum_\mathbf{k}\rightarrow
 \int \rmd^3 k \,\mathcal{V} /(2\pi)^3
\end{math}
and transform the $\bk$-integral to cylindrical coordinates 
$(k,\phi,k_z)$ with the $k_z$-axis aligned to 
$\mathbf{q}$, which yields
\begin{gather}
\nonumber
p_0^{v}(t)=
-\frac{m I_0^2\mathcal{V}}{\pi^2\hbar^2\beta_a}
\sum_{\mathbf{q}\neq0}
q_\perp^2|w_\mathbf{q}|^2
\int_{-\infty}^{\infty}\!\rmd k_z
\log(1-\eta e^{-\beta_a\frac{\hbar^2k_z^2}{2m}})\\
\label{nsum2}
\times 
\frac{\sin^2[\Delta_{q k_z} \frac{t}{2}]}{\Delta_{q k_z}^2}
\end{gather}
with
\begin{math}
 \Delta_{q k_z}\equiv \omega_0+\omega_{q}-{\hbar qk_z}/{m}
\end{math}
and $q\equiv\sqrt{q_\perp^2+q_z^2}$.

For an atomic gas at a temperature $T_a>T_{\text{BEC}}$, 
the fugacity obeys the inequality $\eta<1$. 
Hence, we can Taylor-expand the logarithm in Eq.(\ref{nsum2}) and arrive at
\begin{equation}\label{noft}
p_0^{v}(t)=\frac{m^2 I_0^2 \mathcal{V}}{
\pi^2\hbar^3\beta_a}
\sum\limits_{j=1}^\infty
\frac{\eta^j}{j}
\sum\limits_{\mathbf{q}\neq0}
\frac{q_\perp^2|w_\mathbf{q}|^2}{q}
\mathcal{F}_j(q),
\end{equation}
where we have introduced the convolution integral
\begin{equation}
\label{Fj1}
\mathcal{F}_j(q)\equiv
\int_{-\infty}^{\infty}\rmd\Delta\,
\frac{\sin^2[\Delta \frac{t}{2}]}{\Delta^2} 
e^{-j\frac{\beta_a \hbar}{4\omega_q}(\omega_0+\omega_q-\Delta)^2}.
\end{equation}

Integrals of this type are the essence of Fermi's Golden Rule 
\cite{cohentannoudjiBOOK1}. 
If either the width $q/\sqrt{j\beta_a m}$ of the thermal Gaussian, 
or its central frequency $\omega_0+\omega_q$
are far bigger than the width $2\pi/t$ of the sinc-function
localized at $\Delta=0$, \ie, 
$t\gg \min{(\sqrt{j\beta_a m}/q,2\pi/\omega_0)}$, then we can 
approximate the integral by evaluating the Gaussian function at 
the maximum 
of the sinc-function as
 \begin{align}
\nonumber
\mathcal{F}_j(q)
&\approx
e^{-j\frac{\beta_a \hbar}{4\omega_q}
(\omega_0+\omega_q)^2}
\int_{-\infty}^{\infty}\rmd\Delta\,
 \frac{\sin^2[\Delta \frac{t}{2}]}{\Delta^2}
 \\ \label{Fj2}
&=t \,\frac{\pi}{2}
e^{-j\frac{\beta_a \hbar}{4\omega_q}
(\omega_0+\omega_q)^2}.
\end{align}
This expression displays the familiar linear increase of the 
excitation as a function of time.

\subsection{Relaxation rate}
The evaluation of relaxation rates is always based on time-dependent perturbation theory, either in the Schr\"odinger- or in the Heisenberg picture, and a subsequent evaluation of the amount 
of excitation observed after turning on the interaction for some time. The linear temporal increase of the excitation then translates into a rate coefficient.

\subsubsection{General expression}
Substituting \eq{Fj2} into \eq{noft} and taking the 
time-derivative at the initial instant, 
$\Gamma_0^{v}\equiv\dot{p}_0^{v}(t=0)$, 
we obtain the excitation rate  
\begin{equation}
\label{rate2}
 \Gamma_0^{v}=
\frac{m^2 I_0^2 L}{4\pi^2\hbar^5\beta_a}
\sum\limits_{j=1}^\infty\frac{\eta^j}{j}
\int\limits_{0}^\infty\rmd q\,q^2|V(q)|^2
 e^{-j\frac{\beta_a \hbar}{4\omega_q}
(\omega_0+\omega_q)^2}.
\end{equation} 
We note the reappearance of the two-dimensional 
Fourier-transform of the Casimir-Polder potential 
\eq{Vvonq}, depicted in Fig.~\ref{fig:Vnq}.

In terms of the dimensionless wave number 
$\bar{q}=qR$ of \eq{Vvonq}, we can rewrite \eq{rate2} as
\begin{equation}
\label{rate3} 
\Gamma_0^{v}=\mathcal{A}_0\sum\limits_{j=1}^\infty\frac{\eta^j}{j}\int\limits_{0}^\infty\rmd \bar{q}\,
|V(\bar{q})|^2
\delta_j^{(0)}(\bar{q}).
\end{equation}
Here, we gathered all the constants into the prefactor 
$\mathcal{A}_l\equiv8\pi mI_l^2 L/\hbar^3\lambda_{\text{dB}}^5$
and introduced the sharply peaked function
\begin{equation}
\label{Deltaj}
\delta_j^{(l)}(\bar{q})=\frac{4 \bar{q}^2}{\sqrt{\pi}\varkappa^3}
\exp\left[{-j\left(\frac{\bar{q}}{\varkappa}+
\frac{\varkappa}{\bar{q}}\frac{\hbar \omega_l\beta_a}{4}\right)^2}\right]
\end{equation}
that depends on the characteristic ratio 
\begin{math}
\varkappa
\equiv 
4\sqrt{\pi}{R}/{\lambda_{\text{dB}}}
\end{math} 
of the tube's radius $R$ to the thermal de~Broglie wavelength  
\begin{math}
 \lambda_{\text{dB}}
\equiv
\hbar\sqrt{2\pi\beta_a/m}.
\end{math} 

Indeed, $\delta_j^{(l)}(\bar{q})$
decreases very rapidly as $\bar{q}\rightarrow 0$, as well as for $
\bar{q}\rightarrow \infty$ and we will treat it like a 
delta-distribution
compared to the smooth potential in the integral of \eq{rate3}. 
We can determine its extremum from 
\begin{math}
\partial_{\bar{q}}\delta_j^{(l)}(\bar{q})=0,
\end{math}
and find its location at
\begin{equation} 
\label{qjl}
\bar{q}_{j}^{(l)}=\frac{\varkappa}{\sqrt{2j}}
\{1+[1+\left({j}\hbar\omega_l\beta_a/2\right)^2]^{1/2}\}^{1/2}.
\end{equation}
Therefore, we approximate the integral of \eq{rate3} with the mean value theorem of integral calculus 
\begin{equation}
\label{as_delta}
\int_{0}^\infty\rmd \bar{q}\,
|V(\bar{q})|^2\delta_j^{(0)}(\bar{q})
\approx
|V(\bar{q}_j^{(0)})|^2
\int_{0}^\infty\rmd \bar{q}\,\delta_j^{(0)}(\bar{q}),
\end{equation}
which with the help of the known integral formula \cite{gradsteyn00}
\begin{equation}
\int_{0}^\infty\rmd \bar{q}\,
\delta_j^{(0)}(\bar{q})=
 \frac{e^{-j\hbar\omega_0\beta_a}}{j^{3/2}}
(1+\frac{j}{2}\hbar\omega_0\beta_a),
\end{equation}
finally yields the excitation rate
\begin{equation}
\label{rate4} 
\Gamma_0^{v}=\mathcal{A}_0\sum_{j=1}^\infty
\frac{e^{j\beta_a(\mu-\hbar\omega_0)}}{j^{5/2}}
(1+\frac{j}{2}\hbar\omega_0\beta_a) 
|V(\bar{q}_j^{(0)})|^2\,.
\end{equation}
Clearly, the first $j=1$ term dominates the sum.
\subsubsection{Fermi's Golden Rule}
A simple view of the previous result
comes from considering the unpolarized total transition rate 
\begin{align}
\Gamma_0^v&=\sum_{\bk_f,\sigma_f}\sum_{\bk_i}\,\frac{2\pi}{\hbar}|\bra{f}\hat{H}_i\ket{i}|^2
\delta(\varepsilon_f-\varepsilon_i) n_{\bk_i}
\end{align}
for the inelastic process induced by the Casimir-Polder potential of \eq{hiprime}.
This expression represents the signal current of a fictitious detector that records all 
final states of the atoms and averages it with the thermal distribution of initial states.
Here, the initial state $\ket{i}=\ket{1_{\bk_i},0_{\sigma_i}}$ describes a single atom 
with wave vector $\bk_i$ and the carbon-nano tube in its ground state. 
This state has an energy 
$\varepsilon_i=\frac{\hbar^2}{2m}\bk_i^2$. As we are interested in inelastic collisions, 
we consider the final state 
$\ket{f}=\ket{1_{\bk_f},1_{\sigma_f}}$ with one phonon in the $\sigma$-polarized mode $\omega_0$. Now, this state 
has an energy $\varepsilon_f=\hbar\omega_0+ \frac{\hbar^2}{2m}\bk_f^2$. 
With a few algebraic transformations, we obtain the excitation rate from Fermi's Golden Rule and it coincides with the first term $j=1$ of \eq{rate4}
\begin{equation}
\label{rate5} 
\Gamma_0^{v}=\frac{8\pi m I_0^2 L}{\hbar^3\lambda_{\text{dB}}^5} e^{\beta_a(\mu-\hbar\omega_0)}
(1+\tfrac{1}{2}\hbar\omega_0\beta_a) 
|V(\bar{q}_1^{(0)})|^2.
\end{equation}
In order to bring out the essential physics, one can approximate the fugacity by
$e^{\beta_a\mu}\approx n \lambda_{\text{dB}}^3$ for temperatures $T_a> T_{\text{BEC}}$ (comp. \eq{poly}). This simplifies the rate to
\begin{equation}
\label{rate6} 
\Gamma_0^{v}=\frac{4 m^2 I_0^2 L}{\hbar^5} 
\left(k_B T_a+\tfrac{1}{2}\hbar\omega_0\right) 
\,n 
e^{-\frac{\hbar\omega_0}{k_B T_a}}
|V(\bar{q}_1^{(0)})|^2\,.
\end{equation}
Thus, decreasing the thermal energy of the atoms below the excitation energy, \ie, 
$k_B T_a<\hbar \omega_0$, suppresses the inelastic relaxation rate exponentially..


\subsubsection{Application to current experiment}
In order to evaluate the relaxation rate \eq{rate4} explicitly,
 we need to know the exact form of the Casimir-Polder potential 
of \eq{Vvonq}. Recent measurements
 of the interaction potential between the 
carbon nano-tube and the atomic gas \cite{schneeweiss12} 
have shown that the effect is 
very well described by the contribution $C_5/\rho^{5}$ with a 
numerical coefficient 
$C_5\equiv6\times 10^{-65\pm1} \mbox{Jm}^5$.
Thus, we approximate the Casimir-Polder potential \eq{Vvonq} 
only by that term, \ie,
\begin{align}
V(\bar{q}_1^{(0)})&=2\pi C_5 \frac{V_5(\bar{q}_1^{(0)})}{R^3},
\end{align}
and
\begin{math}
V_5(\bar{q})=
\frac{\bar{q}^{3}}{9}
+\frac{1}{3}\,_1\!{\rm F}\!_2(-\tfrac{3}{2};\{1,-\tfrac{1}{2}\};-\tfrac{\bar{q}^{2}}{4})
\approx \frac{1}{3}-\frac{\bar{q}^2}{4}+\mathcal{O}[\bar{q}^3].
\end{math}
As the $R\ll \lambda_{\text{dB}}$, we need to retain only lowest term in this expression and find the following approximation for the rate
\begin{equation}
\label{rate7} 
\Gamma_0^{v}=\frac{16 \pi^2 m^2 I_0^2 L}{9 \hbar^5} 
\left(k_B T_a+\tfrac{1}{2}\hbar\omega_0\right) 
\,n 
e^{-\frac{\hbar\omega_0}{k_B T_a}}
\frac{C_5^2}{R^6}.
\end{equation}
All other relevant parameters of the carbon nano-tube and the 
atomic gas are listed in Apps.~\ref{normalization} and
\ref{app:fugacity}.

\begin{figure}[h]
\includegraphics[clip,width=\columnwidth]{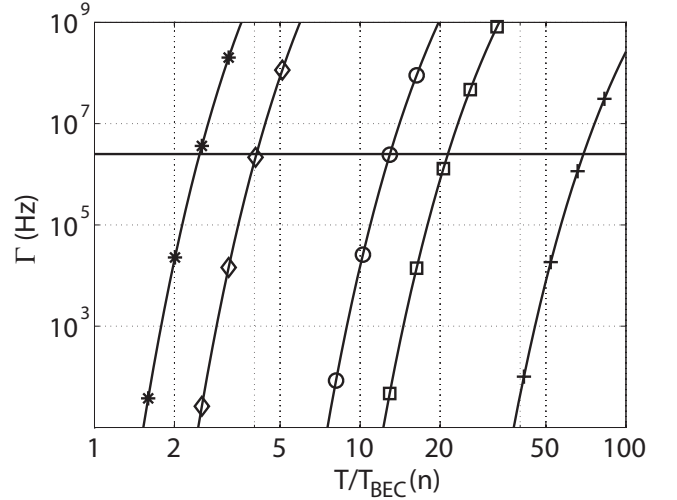}
\caption{\label{Rate_Bild2} 
Excitation rate $\Gamma_0^{v}$ versus scaled 
temperature $T/T_{\text{BEC}(n)}$ 
for different densities $n$ of the atomic cloud:
$n=10^{12} \mbox{cm}^{-3}$ ($+$),
 $5\cdot10^{12}\mbox{cm}^{-3}$ ($\square$),
 $10^{13} \mbox{cm}^{-3}$ ($\circ$),
  $5\cdot10^{13}\mbox{cm}^{-3}$ ($\Diamond$),
  $10^{14} \mbox{cm}^{-3}$ ($*$).
The horizontal line marks the frequency 
$\omega_0=2\pi\cdot 398$~kHz of the lowest phononic mode 
and represents the boundary between
the over-damped (above) and the under-damped (below) oscillations of the carbon nano-tube.} 
\end{figure}

Now, figure~\ref{Rate_Bild2} depicts the dependence of the full excitation
rate $\Gamma_0^{v}$ of \eq{rate4}
on the scaled temperature $T_a/T_{\text{BEC}}$ of the atomic cloud for different atomic densities. 
We note that the excitation rate depends strongly on the 
temperature, as well as on the density of the gas. 
For the given parameters, we find the rates of \eq{rate4} and \eq{rate7} are indistinguishable.
When the 
excitation rates are of the order of the oscillation frequency or 
below $\Gamma_0^{v}<\omega_0$, the Fermi-Golden-Rule approach is 
suitable. Therefore, we conclude that cooling of the carbon
 nano-tube to the ground mode is
 feasible in current experimental situations.

\section{Dynamics of  relaxation}
\label{DensM}
In this section, we will generalize the previous considerations 
to determine the relevant time scales and study the dynamics of 
the relaxation process with a master equation approach. 
Here, the phonons are considered as the system and the atoms are the bath. This analysis yields (i) a dynamic picture of the carbon nano-tube  cooling, (ii) its approach towards equilibrium, and (iii) its dependence on  temperature.

\subsection{Master equation}

In the interaction picture, the time evolution of the total 
density operator $\hat{\tilde{\rho}}_{\text{tot}}$ of phonons and atoms is described by the von-Neumann equation 
\begin{equation}
\label{vonNeumann}
\dot{\hat{\tilde{\rho}}}_{\text{tot}}(t)=\frac{1}{i \hbar}\left[{\hat{\tilde{H}}_{i}(t)},{\hat{\tilde{\rho}}_{\text{tot}}(t)}\right].
\end{equation} 
As usual, the interaction picture 
is obtained from the Schr\"odinger picture using 
the free atom-phonon propagator 
$\hat{U}(t)\equiv \exp[-i(\hat{H}_{\text{a}}+\hat{H}_{\text{c}})t/\hbar]$, which yields
\begin{align}
\hat{\tilde{\rho}}_\text{tot}(t)&\equiv
\hat{U}^\dagger(t)\hat{\rho}_\text{tot}(t)\hat{U}(t),\\
\label{hiint}
\hat{\tilde{H}}_i(t)&\equiv \hat{U}^\dagger(t)\hat{H}_{\text{i}}
\hat{U}(t)=
\hbar \sum_{\vec{q}\neq 0}w_\bq\hat{\tilde{\mathcal{N}}}_\bq(t)
\hat{\tilde{\mathcal{T}}}_\bq(t),\\
\hat{\tilde{\mathcal{N}}}_\bq(t)&=\sum_{\bk}
e^{-i(\omega_{\bk}-\omega_{\bk-\bq})t} 
\,\aopd{\vec{k}-\vec{q}}\aop{\vec{k}},\\
\label{tint}
\hat{\tilde{\mathcal{T}}}_\bq(t)&=\int\limits_0^L \frac{\D z}{L}\,
\exp{[\I \bq \hat{\tilde{\mathbf{u}}}(z,t)]},
\end{align}
and
\begin{align}
\label{u_quant1}
\hat{\tilde{\mathbf{u}}}(z,t)&=
\frac{1}{\sqrt{2}}\sum_{\substack{l=0\\
\sigma=x,y}}^{\infty}
\mathbf{e}_{\sigma}\phi_l(z)(e^{-i\omega_l t}\hat{b}_{l\sigma}^{\phantom{\dagger}}+
e^{i\omega_l t}\hat{b}_{l\sigma}^{\dagger}).
\end{align}

To study the phonon dynamics 
$\hat{\tilde{\rho}}(t)\equiv \text{Tr}_a[\hat{\tilde{\rho}}_{
\text{tot}}(t)]$, we have to average the state of total system
over the atomic bath. Considering the deviation 
$\Delta\hat{\tilde{\rho}}(t)\equiv\hat{\tilde{\rho}}(t+\Delta t)-\hat{\tilde{\rho}}(t)$
for a short time interval $\Delta t$, we find
from an iterated formal solution of \eq{vonNeumann} the expression
\begin{gather}
\label{eq:master_init}
\Delta\hat{\tilde{\rho}}(t)=
\frac{1}{i\hbar}\int_t^{t+\Delta t}
\D t_1\,\traceind{a}{\kommuthat{\hat{\tilde{H}}_i(t_1)}{
\hat{\tilde{\rho}}_{\text{tot}}(t)}}\\ 
-\frac{1}{\hbar^2}\int_t^{t+\Delta t}\!\D t_1\int_t^{t_1}\!\D t_2
\traceind{a}{\kommuthat{\hat{\tilde{H}}_i(t_1)}{
\kommuthat{\hat{\tilde{H}}_i(t_2)}{\hat{\tilde{\rho}}_{\text{tot}}(t_2)}}}.
\nonumber
\end{gather}
Within the Born-Markov approximation \cite{cohentannoudjiBOOK1},
we derive a master equation for the coarse-grained 
rate of change $\Delta\hat{\tilde{\rho}}(t)
\approx\Delta t \frac{d\hat{\tilde{\rho}}}{dt}$.
For the particular atom-phonon interaction Hamiltonian given by \eq{hiint}, 
we obtain the master equation 
\begin{gather} 
\label{eq:master_before_energy}
\frac{d\hat{\tilde{\rho}}}{dt}=
\frac{\eta \mathcal{V} }{\lambda_\text{dB}^3}
\int_0^\infty\frac{\D\tau}{\Delta t}\!
\int_t^{t+\Delta t}\D  t'
\sum_{\bq\neq 0}
|w_{\bq}|^2
 e^{-i\omega_{\bq}\tau(1-i\frac{\tau}{\hbar\beta})}\\ 
\times
[\tTdq{t'-\tau}\hat{\tilde{\rho}}(t)\tTq{t'}
-\tTq{t'}\tTdq{t'-\tau}\hat{\tilde{\rho}}(t)]+\text{h.c.}.\nonumber
\end{gather} 
In progressing from Eqs.~(\ref{eq:master_init}) to 
(\ref{eq:master_before_energy}), we have not included 
linear contributions in the interaction as there is no deterministic motion.
Moreover, in evaluating thermal contributions for $T_a>T_{\text{BEC}}$, we have retained only the dominant contributions and encountered the same 
thermal density-fluctuation correlation function
$\av{\hat{\tilde{\mathcal{N}}}_{\bq}(t_1) \hat{\tilde{\mathcal{N}}}_{\bk} (t_2)}$,
as in \eq{ptt} and discussed in App.~\ref{ttc}.
Since the thermal correlation functions also decay quickly, it is 
very well justified to extend the upper limit of the 
$\tau$-integration to infinity.

\subsection{Ground state excitation rate}
\label{sec:groundstate_excitations}
As an application of this master equation, we consider a carbon
nano-tube that is initially in the multi-mode phononic 
vacuum
\begin{math}
\hat{\tilde{\rho}}(t=0)\equiv \ket{0,\ldots}\bra{0,\ldots}.
\end{math}
Then, we immerse it into the bath of atoms at temperature $T_a$,
and observe the decrease of the ground-state occupation 
$\rho^{v}(t)\equiv\bra{0}\hat{\tilde{\rho}}(t)\ket{0}$.
In this case, the excitation rate into any available phononic mode $l\ge 0$ is given by
\begin{align}
\Gamma^{v}&=
  \label{rhopunkt00}
\frac{\eta \mathcal{V}}{\lambda_\text{dB}^3}
\int_0^\infty
\frac{\D\tau}{\Delta t}\int_0^{\Delta t}\D  t'
\sum_{\vec{q}\neq 0}
|w_{\vec{q}}|^2
 e^{-i\omega_{\bq}\tau(1-i\frac{\tau}{\hbar\beta_a})}\nonumber\\
&\times
\bra{0}\hat{Q}_\bq(t',t'-\tau)\ket{0}
  +\text{h.c.}.\\
\hat{Q}_\bq(t_1,t_2)&\equiv
	\tTq{t_1}\tTdq{t_2}
-\tTdq{t_2}\ket{0}\bra{0}\tTq{t_1}.
	\end{align}
Clearly, the excitation rate is just the negative depletion rate of the ground-state occupation, \ie,
$\Gamma^{v}\equiv -\dot{\rho}^{v}(t)$.

As in Sec.~\ref{sectoteng}, we assume  that the thermal 
de~Broglie wavelength of the atomic gas is much longer than the tube's oscillation amplitude, that is
$\lambda_\text{dB}\gg u$. Hence, we can approximate the 
translation operator of \eq{tint}, as in \eq{Texpanded}, by a 
second-order Taylor series, and find
\begin{gather} 
\bra{0}\hat{Q}_\bq(t',t'-\tau)\ket{0}
=
\sum_{\substack{l=0\\\sigma=x,y}}^{\infty}(\vec{qe}_\sigma)^2I_l^2
e^{-i \omega_l \tau}.
\end{gather} 
After performing the time integrations and 
transforming the discrete wave number sums in the continuum limit, we arrive for the vacuum excitation rate at
\begin{equation}
\label{eq:gamma00int}
\Gamma^{v}=\sum_{l=0}^{\infty}\!\mathcal{A}_l\eta\int_{0}^\infty\rmd \bar{q}\,
|V(\bar{q})|^2
\delta_1^{(l)}(\bar{q}).
\end{equation}
The $\bar{q}$-integration is approximated in a way completely analogous 
to \eq{as_delta} and we obtain the expression 
\begin{gather}
\label{eq:exci_sum}
\Gamma^{v}=\sum_{l=0}^{\infty} \mathcal{A}_l
e^{\beta_a(\mu-\hbar\omega_l)}
(1+\frac{1}{2}\beta_a\hbar\omega_l)
|{V}(\bar{q}_1^{(l)})|^2
\end{gather}
for the phononic excitation rate out of the vacuum. Clearly, the sum in \eq{eq:exci_sum} is dominated by the contribution of the lowest phononic mode $l=0$, that is
\begin{equation}
\label{eq:excitation_V_allg}
\Gamma^{v}_0
\approx 
\mathcal{A}_0 
e^{\beta_a(\mu-\hbar\omega_0)}
(1+\frac{1}{2}\beta_a\hbar\omega_0)
|V(\bar{q}_1^{(0)})|^2.
\end{equation}  
This expression for the rate agrees with 
the first term of the extended thermal series \eq{rate4}.

\subsection{Finite temperature thermalization rate}
\label{sec:thermal_damping_rates}
In this subsection, we generalize the previous calculation by
 assuming 
that the carbon nano-tube is  initially close to a thermal state at temperature $T_c\equiv 1/k_B\beta_c$, 
which is different from the temperature $T_a\equiv 1/k_B \beta_a$
 of the atomic bath.   
The rate of change of the occupation 
${p}_{l,\sigma}
\equiv
\erwart{\hat{p}_{l,\sigma}}=
\erwart{\hat{b}_{l,\sigma}^\dagger 
\hat{b}_{l,\sigma}^{\phantom\dagger}}$
of the mode $(l,\sigma)$ is given by
\begin{equation}
\label{eq:start:thermal_damping}
\dot{p}_{l,\sigma}=
\trace{\hat{p}_{l,\sigma}\dot{\hat{\tilde{\rho}}}(t)}.
\end{equation}

Now, we use the master equation \eq{eq:master_before_energy} and 
assume in the evaluation of the averages a canonical density 
operator parametrized by a time-dependent 
coefficient $\beta_c(t)$, as outlined in App.~\ref{phononcanon}. 
In the continuum-limit, we obtain from \eq{eq:start:thermal_damping} for the unpolarized occupation numbers 
\begin{math}
p_l\equiv {p}_{l,x}+{p}_{l,y}
\end{math}
of the l-{th} level the rate equation
\begin{align}
\label{eq:thermal_damping_discrete}
\dot{p}_l(t)&=-\gamma_l(\beta_c)p_l,
\end{align}
with temperature dependent rate coefficients
\begin{align}
\gamma_{l}(\beta_c)&\equiv 
\mathcal{A}_l\eta
(e^{\beta_a\hbar\omega_l}-e^{\beta_c\hbar\omega_l})
\int_{0}^\infty\rmd \bar{q}\,
|V(\bar{q})|^2\delta_1^{(l)}(\bar{q}).
\end{align} 
Integrating over $\bar{q}$  similarly to  
Eqs.~(\ref{rate3}) and (\ref{as_delta}), we find
\begin{equation}
\label{eq:gamma_thermal}
\gamma_l=\mathcal{A}_l\eta
(1-\e{(\beta_c-\beta_a)\hbar\omega_l}) (1+\frac{\beta_a\hbar\omega_l}{2})
|{V}(\bar{q}_1^{(l)})|^2,
\end{equation}
with $\bar{q}_1^{(l)}$ given by \eq{qjl}.

As expected, this expression generalizes Eqs.~(\ref{rate4}) and (\ref{eq:exci_sum}), as 
it allows for a finite temperature of the tube that is different from the atomic gas, \ie, $\beta_c\neq\beta_a$, as well as for 
excitations into all phononic modes $l\ge 0$. 

If the carbon nano-tube is hotter than the atomic bath, that is
$\beta_c<\beta_a$,
then the relaxation rate of \eq{eq:gamma_thermal} is positive 
and leads, according to 
\eq{eq:thermal_damping_discrete}, to a cooling of the tube's phonons.
If the carbon nano-tube is colder than the atoms, \ie, $\beta_c>\beta_a$, 
the rate is negative and leads to a heating of the tube.
Finally, all rates vanish, when $\beta_c=\beta_a$, as required for a 
thermodynamic equilibrium.

\section{Conclusion}
Using time-dependent perturbation theory with finite temperature ensembles, 
we have calculated the excitation rate of a free-standing single-walled carbon nano-tube immersed in a bath of neutral bosonic atoms.
The interaction between the carbon nano-tube and the atoms was modeled by a generic Casimir-Polder potential series. We have assumed that the temperature of the atoms was above the Bose-Einstein condensation transition temperature as the collisional relaxation at the MHz level is insensitive to the phase coherence of the bath. 

For the numerical evaluation of the excitation rates, we have used experimentally determined  values of the Casimir-Polder potential between a thermal-, as well as  a Bose-Einstein condensed $^{87}$Rb gas \cite{schneeweiss12}. In this situation, an inverse power-law $\sim C_5/r^{5}$ is very well-suited to approximate this potential.  With this analysis and the current data, we find that cooling of the free-standing carbon nano-tube to the phononic ground state  due to the Casimir-Polder interaction with a cold atomic gas is feasible.

We emphasize that the excitation rate depends strongly on the temperature and the density of the atomic cloud. The form of 
the interaction potential, for example interferences between different potential contributions, can influence the rate as well. Hence, more extensive experimental data is needed.

\section*{Acknowledgements}
JF and CTW gratefully acknowledge support from the German BMBF 
(NanoFutur 03X5506). WPS and JF cooperated within 
the SFB/TRR 21 {\em ``Control of quantum correlations in tailored matter''} funded by the Deutsche Forschungsgemeinschaft (DFG), and RW thanks the Deutsche Luft- und Raumfahrtagentur (DLR) 
for support from grant (50WM 1137).

\appendix
\section{Properties of the carbon nano-tube
\label{normalization}}
In this appendix, we summarize without much of a derivation properties of the carbon nano-tube relevant for the article.

The normalization constant $\tilde{a}_l$ of the eigenmode of \eq{eigenmode} reads
\begin{align} 
\label{eqnorm}
\frac{\tilde{a}_l^2}{a_l^2}&=
1+2\cos{(\kappa_lL)}\cosh{(\kappa_lL)}
+\frac{1}{2}[\cos{(2 \kappa_lL)}\\
&+\cosh{(2\kappa_lL)}]
-\frac{1}{2\kappa_lL}
[2\cosh{(\kappa_lL)}\sin{(\kappa_lL)}\nonumber\\
&+2\cos{(\kappa_lL)}\sinh{(\kappa_lL)}
+\cosh^2{(\kappa_lL)}\sin{(2\kappa_lL)}\nonumber\\
&+\cos^2{(\kappa_lL)}\sinh{(2\kappa_lL)}].\nonumber
\end{align}

Typical mechanical parameters for single-walled carbon nano-tube are summarized in Tab.~\ref{tab1}.
\begin{table}[h]
\begin{ruledtabular}
\begin{tabular}{c c c c c} 
 $R$ [nm]&  $L$ [$\mu$m]&  $\rho_c$ [kg/m] &  $\omega_0$ [kHz]& $a_0$ [nm]\\ \hline
 $1$ &  $1$ &  $10^{-15}$  & $2\pi\cdot 398$ & 0.2 
\end{tabular}
\end{ruledtabular}
\caption{\label{tab1}Mechanical parameters of a single-walled carbon nano-tube.}
\end{table}

In the low temperature approximation of \eq{Texpanded}, we 
have assumed that the spatial excursions of the nano-tube are much 
less than the thermal de~Broglie wavelength of the atomic gas. 
Thus, we summarize \cite{weissbook10} the maximal thermal excursions 
$u=\sqrt{\erwart{\Delta^2 \hat{u}}_{T_c}}$ of the tip of the 
tube $z=L$ for 
different temperatures of the tube $T_c$ in Tab.~\ref{tab2} 
\begin{table}[h] 
\begin{ruledtabular}
\begin{tabular}{rccc}
$T_c$ [K] & 4 & 0.24 &0 \\ \hline 
$u$ [nm] & 270 & 66 & 0.46
\end{tabular}
\end{ruledtabular}
\caption{\label{tab2}
Maximal thermal displacement $u$ at the tip of the 
carbon nano-tube for three values of temperature $T_c$.}
\end{table}
\section{Cold bosonic gases\label{app:fugacity}}
In this appendix, we summarize the central results of equilibrium thermodynamics of bosonic fields that were employed in the main sections of our article.
\subsection{Bose-Einstein condensation}
The state of the free atomic gas \cite{stringaribuch03} is 
described by the density operator of the grand canonical ensemble 
\begin{equation}
\hat\rho=\e{{\Omega}-\beta_a(\hat{H}-\mu\hat{N})},\quad \trace{\hat\rho}=1,
\end{equation}
held at a temperature $T_a\equiv 1/k_B \beta_a$ and maintaining an average particle number $N$. The Hamiltonian is denoted by 
$\hat{H}$, $\Omega$ is the grand canonical potential and 
$\mu$ is a chemical potential, which is determined 
self-consistently through the particle number constraint 
\begin{equation}
\label{totnum}
N\equiv\erwart{\hat{N}}=
\sum_{\vec{k}}n_{\vec{k}}=n_0+\tilde{N}(\beta_a).
\end{equation}
Here $n_0\equiv n_{\bk=0}$ is the occupation of the ground-state and 
$\tilde{N}(\beta_a)$ denotes the number of thermal particles.
 
Introducing the fugacity as $\eta\equiv e^{\beta_a\mu}$, 
one finds for the occupation number of the ground state, 
$n_0\equiv\eta/(1-\eta)$ and 
$\tilde{N}(\beta_a)=\mathcal{V} 
g_{3/2}(\eta)/{\lambda_\text{dB}^3}$ for 
the number of particles in the excited states. 
The poly-logarithmic function is defined by
\begin{equation}
\label{poly}
g_{3/2}(\eta)\equiv\sum_{j=1}^\infty \frac{\eta^j}{j^{3/2}}, \qquad g_{3/2}(1)\approx 2.61238.
\end{equation} 

In the thermodynamical limit, when particle number and volume 
increase at constant particle density, \ie, $n=\lim_{N,\mathcal{V}
\rightarrow\infty}N/\mathcal{V}=\text{const.}$,
one can identify two cases. 
The first corresponds to particle densities below the critical 
density, that is, $n<g_{3/2}(1)/\lambda_\text{dB}^3$. 
There, the  occupation number of the ground state $n_0$ is 
negligibly small  and  the total number of atoms is equal to the 
number of particles in the excited states
\begin{math}
N\approx \tilde{N}(\beta_a).
\end{math} 
In this limit of a dilute quantum gas, 
the de~Broglie wavelength is much smaller than the average 
distance between the particles. Therefore, the gas behaves classically.

The second regime occurs for densities larger than 
the critical density $n>g_{3/2}(1)/\lambda_\text{dB}^3$.
In this case the fugacity attains its maximum value 
$\eta\rightarrow 1$. 
The occupation of the ground state 
\begin{equation}
n_0=N\left[1-\left(\frac{T_a}{T_{\text{BEC}}}\right)^\frac{3}{2}\right],
\end{equation}
is not negligible anymore and 
increases with the decrease of temperature
where we have introduced the critical BEC temperature
\begin{equation}
\label{Tc}
T_{\text{BEC}}\equiv
\frac{2\pi\hbar^2}{mk_B}\left[\frac{n}{g_{3/2}(1)}\right]^{2/3}.
\end{equation}

\begin{figure}[htbp]
\centering\includegraphics[clip,width=\columnwidth]{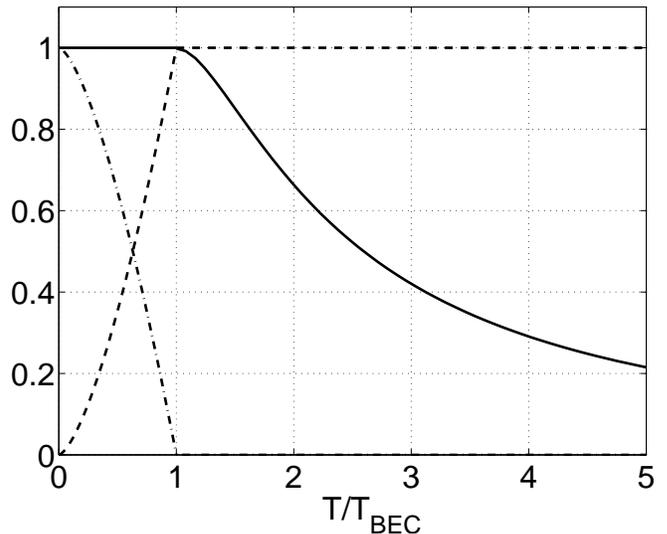}
\caption{Fugacity $\eta$ as a function of scaled
temperature $T/T_{\text{BEC}}$ (solid line), fraction of particles in the ground-state  $n_0/N$ (dashed-dotted line), and 
fraction of particles in the excited states $\tilde{N}/N$ 
(dashed line).}
\label{fig:BECandFugaz}
\end{figure}

Figure~\ref{fig:BECandFugaz} presents the behavior of the gas in
 these two cases. For temperatures below the critical 
temperature, the ground-state occupation
(dashed-dotted line) becomes macroscopic and the fugacity (solid line) 
is practically one.  Above the critical temperature, the 
fugacity decreases with the increase of temperature and the 
number of particles in the excited states (dashed line) 
is equal to the total number of particles.

\begin{table}[hT] 
\begin{ruledtabular}
\begin{tabular}{rcccccc}
$n$  [cm$^{-3}$] & $10^{12}$ &$5\cdot10^{12}$ & $10^{13}$ & $5\cdot10^{13}$&
$10^{14}$  \\\hline
$T_\text{BEC}$ [nK] & 18 & 54& 85& 250 & 400\\\hline
$\lambda_\text{dB}$ [nm] & 610&  357&  283 & 165 &131 
\end{tabular}
\end{ruledtabular}
\caption{\label{tab3} 
Parameters of the cold atomic cloud. The mass of a single $^{87}{\rm Rb}$ atom is  $m=1.443\cdot10^{-25}\:\mbox{kg}$.}
\end{table}

From this analysis, we can obtain the values of the thermal 
de~Broglie wavelength, summarized in Tab.~\ref{tab3}, for a range of atomic densities at the
critical temperature.

\subsection{Thermal correlation functions}
\label{ttc}
In this part of the appendix we focus on the quantum averages 
used in the master equation~\eq{eq:master_before_energy}. 
In particular, we derive single-time averages and two-time correlation functions.

\subsubsection{Single-time average}
With the choice of the interaction Hamiltonian \eq{hiint},
and assuming that the atomic bath is in a grand-canonical state without macroscopic motion, we find a vanishing quantum average
\begin{equation}
\traceind{a}{\hat{\tilde{\rho}}_a\hat{\tilde{H}}_i}=0,
\end{equation} 
because $\erwart{\aopd{\vec{k}-\vec{q}}\aop{\vec{k}}}=0$, for $
\vec{q}\neq 0$. Therefore, the trace over the single commutator in the first 
term of \eq{eq:master_init}, will vanish -- which means 
that we have incorporated this energy shift in the unperturbed 
Hamiltonian of the atoms. This result is well-known in 
first-order time-independent perturbation theory.

\subsubsection{Two-time correlations}
The two-time density-fluctuation correlation function
 $\av{\hat{\tilde{\mathcal{N}}}_\bq(t_1)
\hat{\tilde{\mathcal{N}}}_{\bk}(t_2)}$ evaluated 
at temperature $T_a$,
emerges ubiquitously 
in field theory and is proportional to the 
dynamic structure factor \cite{pines63}. 
In the case of stationary, 
translationally invariant systems, it simplifies to
\begin{gather}
\nonumber
\av{\hat{\tilde{\mathcal{N}}}_\bq(t+\tau)
\hat{\tilde{\mathcal{N}}}_{\bk}(t)}_=
\av{\hat{\tilde{N}}_\bq(\tau)
\hat{\tilde{N}}_{\bk}(0)}
=\delta_{\bq+\bk,0}
\,g\pd_{\vec{q}}(\tau),
\end{gather}
where the Kronecker delta enforces momentum conservation 
and we find the thermal correlation function
\begin{gather}
\label{eq:corr_int}
g\pd_{\vec{q}}(\tau)\equiv
\sum_{\vec{k}}e^{\I(\omega_{\bk-\bq}-\omega_{\bk})\tau}
n_{\bk-\bq}(1+n_{\bk})
\end{gather} 
with thermal occupations $n_{\vec{k}}$ defined in \eq{GrCan}. 

For further discussions, we separate the ground-state contribution 
from the sum and approximate the remainder within the continuum-limit. For $T_a> T_\text{BEC}$, we find $n_{\vec{k}}<1$ and 
we will take only the linear term  in~\eq{eq:corr_int} into 
account. In this regime, we can also approximate the Bose-Einstein 
distribution by the Maxwell-Boltzmann distribution. Moreover, the 
occupation number in the lowest level is negligible and the 
correlation function can be written as
\begin{equation}
\label{eq:corr_ana}
g_{\vec{q}}(\tau)\approx \frac{\eta\mathcal{V}}{
\lambda_\text{dB}^3}\exp{[-i\omega_{\bq}\tau(1-i\frac{\tau}{
\hbar\beta_a})]}.
\end{equation} 

\subsection{Canonical ensemble for phonons}
\label{phononcanon}
In this subsection, we recall general properties of the 
canonical ensemble 
\begin{equation}\label{rhofinApp}
\hat{\rho}_c\equiv
e^{\bar{\Omega}-\beta_c\hat{H}_c}, \quad 
\text{Tr}_c{\{\hat{\rho}_c\}}=1,
\end{equation} 
 which is used when the system exchanges energy but no particles 
with the environment. Here, $\bar{\Omega}$ is a canonical potential, $\hat{H}_c$ is the Hamiltonian defined by \eq{hc} for the phonons of the carbon nano-tube,
and $T_c\equiv1/k_B \beta_c$ is the temperature of the phonons.

Therefore, the occupation of the mode $(l,\sigma)$ reads
\begin{gather}
\label{n_av}
p_{l,\sigma}(\beta_c)\equiv
\erwart{\hat{p}_{l,\sigma}}=
\frac{1}{e^{\beta_c\hbar\omega_l}-1},
\end{gather}
and for the density-correlation, we find
\begin{gather}
\label{nn_av}
\erwart{\hat{p}_{l,\sigma}\hat{p}_{l',\sigma'}}
\equiv (1+\delta_{l,l'}\delta_{\sigma,\sigma'}
e^{\beta_c\hbar\omega_l})
p_{l,\sigma} p_{l',\sigma'}.
\end{gather} 
Recalling \eq{eq:master_before_energy}, we obtain from \eq{n_av}
the rate of change 
\begin{align} 
\nonumber
\dot{p}_{l,\sigma}
&=\int_0^\infty{}
\frac{\D\tau}{\hbar^2\Delta t}\int_t^{t+\Delta t}\D t'\sum_{\vec{q}\neq 0}\Big[g\pd_{\vec{q}}(\tau)
\Big(\av{\hat{\tilde{T}}_\bq(t')\hat{p}_{l,\sigma}\hat{\tilde{T}}_\bq^\dagger(t'-\tau)}
\\ 
&-\av{\hat{p}_{l,\sigma}\hat{\tilde{T}}_\bq(t')\hat{\tilde{T}}_\bq^\dagger(t'-\tau)}
\Big) + \text{h.c.}\Big]
\end{align}
of the occupation of the mode $(l,\sigma)$.

With the quadratic approximation of the tube's translation
 operator \eq{Texpanded}, we find
\begin{gather} \nonumber
\av{\hat{p}_{l,\sigma}\hat{\tilde{T}}_\bq(t')
\hat{\tilde{T}}_\bq^\dagger(t'-\tau)}
-\av{\hat{\tilde{T}}_\bq(t')\hat{p}_{l,\sigma}
\hat{\tilde{T}}_\bq^\dagger(t'-\tau)}=
\sum_{l',\sigma'}I_{l'}^2\\ \nonumber
\times(\vec{qe}_{\sigma'})^2\Big[\erwart{\hat{p}_{l,\sigma}}
\left(e^{-\I\omega_{l'}\tau}
-e^{-\omega_{l'}(\beta_c\hbar-\I\tau)} \right)
+\erwart{\hat{p}_{l',\sigma'}\hat{p}_{l,\sigma}}\\
\times\left(2\cos{(\omega_{l'}\tau)}-
e^{\omega_{l'}(\beta_c\hbar-\I\tau)}
-e^{-\omega_{l'}(\beta_c\hbar-\I\tau)}\right)\Big].
\end{gather} 

We note that these expressions are very similar to those 
considered in Sec.~\ref{sec:groundstate_excitations} for the 
ground-state excitation.
Using Eqs.~(\ref{n_av})-(\ref{nn_av}) and integrating over 
$t'$ and $\tau$, we obtain
\begin{align}
\nonumber
\dot{p}_{l,\sigma}(\beta_c)
&=
\frac{\eta m \mathcal{V}}{\hbar\lambda_\text{dB}^2}
\sum_{\vec{q}\neq 0} \frac{|w_{\vec{q}}|^2}{|\vec{q}|}
\sum_{l',\sigma'}I_{l'}^2(\vec{qe}_{\sigma'})^2 
e^{-\frac{\hbar\beta_a(\omega_\bq-\omega_{l'})^2}{4\omega_\bq}}\\
 \nonumber 
 &\times\left( e^{-\beta_c\hbar\omega_{l'}}
-e^{-\beta_a\hbar\omega_{l'}}\right) 
\Big[1+(1+\delta_{l,l'}\delta_{\sigma,\sigma'}
e^{\beta_c\hbar\omega_{l'}})\\ 
\label{thdamp} 
&\times\left(1-e^{\beta_c\hbar\omega_{l'}}\right)
p_{l',\sigma'}(\beta_c)\Big]
p_{l,\sigma}(\beta_c).
\end{align} 
With the help of \eq{n_av}, \eq{thdamp} reduces to
\begin{align}
\nonumber
\dot{p}_{l,\sigma}(\beta_c)
&=\frac{\eta m \mathcal{V}}{\hbar\lambda_\text{dB}^2}
\sum_{\vec{q}\neq 0}\frac{|w_{\vec{q}}|^2}{|\vec{q}|}
I_{l}^2(\vec{qe}_{\sigma})^2
e^{-\frac{\hbar\beta_a(\omega_\bq-\omega_{l})^2}{4\omega_\bq}}\\ 
&\times
\left(e^{(\beta_c-\beta_a)\hbar\omega_{l}}-1\right)
p_{l,\sigma}(\beta_c).
\end{align} 
These technical steps are necessary to obtain the 
finite temperature thermalization rate of \eq{eq:gamma_thermal}.
\bibliography{bec,CNT,textbooks,MyPublications} 
\end{document}

%% file: macro.inc
\newcommand{\rmd}{{\rm d}}

\newcommand{\ie}{i.\thinspace{}e.}

\newcommand{\av}[1]{\langle #1 \rangle}

\newcommand{\eq}[1]{Eq.\;(\ref{#1})}

\makeatletter
\let\@hbar\hbar
\def\math@bold{bold}
\def\hbar{\ifx\math@version\math@bold\hbar@fett\else\@hbar\fi}
\def\hbar@fett{\mathchar'26\mkern-9muh}
\makeatother

 \newcommand{\bra}[1]{\langle{#1}|}
 \newcommand{\ket}[1]{|{#1}\rangle}
 \newcommand{\braket}[2]{\langle{#1}|{#2}\rangle}

 \newcommand{\trace}[1]{\operatorname{Tr}\left\{ #1 \right\}}
 \newcommand{\traceind}[2]{\operatorname{Tr}_{#1}\left\{ #2 \right\}}

 \newcommand{\erwart}[1]{\langle{#1}\rangle}

 \newcommand{\tTq}[1]{\hat{\tilde{T}}\pd_{\vec{q}}(#1)}

 \newcommand{\tTdq}[1]{\hat{\tilde{T}}^\dagger_{\vec{q}}(#1)}

 \newcommand{\e}[1]{\operatorname{e}^{#1}}

 \newcommand{\I}{\text{i}}

 \newcommand{\D}{\text{d}}

 \newcommand{\pd}{^{\phantom\dagger}}

 \newcommand{\aop}[1]{\hat{a}_{#1}^{\phantom \dagger}}
 \newcommand{\aopd}[1]{\hat{a}_{#1}^{\dagger}}

\newcommand{\bk}{\vec{k}}
\newcommand{\br}{\vec{r}}

 \newcommand{\kommuthat}[2]{\left[#1,#2\right]}

\usepackage{ifthen} 
\newcommand{\bu}[1][]{
	\ifthenelse{\equal{#1}{}}
	{\vec{u}}
	{\vec{u}_{#1}}
}
\newcommand{\bv}[1][]{
	\ifthenelse{\equal{#1}{}}
 	{\vec{v}}
	{\vec{v}_{#1}}
} 
\newcommand{\bq}{{\vec{q}}}

\def\vec#1{\ensuremath{\mathchoice{\mbox{\boldmath$\displaystyle#1$}}
                            {\mbox{\boldmath$\textstyle#1$}}
                            {\mbox{\boldmath$\scriptstyle#1$}}
                            {\mbox{\boldmath$\scriptscriptstyle#1$}}}}
\newcommand{\btu}[1][]{
	\ifthenelse{\equal{#1}{}}
	{\tilde{\vec{u}}}	
	{\tilde{\vec{u}}_{#1}}
}
 
\newcommand{\btv}[1][]{
	\ifthenelse{\equal{#1}{}}
	{\tilde{\vec{v}}}
	{\tilde{\vec{v}}_{#1}}
}

 \newcommand{\mycaption}[2]{\caption[#1]{#1\ifx\end#2\end\else . #2\fi}} 